\documentclass{phbauth}        
\usepackage{graphicx}

\begin{document}

\begin{frontmatter}

\title{Proximity effect in planar Superconductor/Semiconductor junction}

\author[address1]{Fran\c cois Lefloch\thanksref{thank1}},
\author[address1]{David Quirion},
\author[address1]{Marc Sanquer}

\address[address1]{DRFMC - SPSMS - CEA/Grenoble, 17 Avenue des Martyrs, 38054 Grenoble cedex 09}

\thanks[thank1]{Corresponding author. E-mail: flefloch@cea.fr} 

\begin{abstract}
We have measured the very low temperature (down to $30\,mK$) subgap resistance of Titanium Nitride (Superconductor, $T_{c}=4.6\,K$)/highly doped Silicon (Semiconductor) SIN junction (the insulating layer I stands for the Schottky barrier). As the temperature is lowered below the gap, the resistance increases as expected in SIN junction. Around $300\,mK$, the resistance shows a maximum and decreases at lower temperature. This observed behavior is due to coherent backscattering towards the interface by disorder in the Silicon (``Reflectionless tunneling''). This effect is also observed in the voltage dependence of the resistance (Zero Bias Anomaly) at low temperature ($T<300mK$). The overall resistance behavior (in both its temperature and voltage dependence) is compared to existing theories and values for the depairing rate, the barrier resistance and the effective carrier temperature are extracted.
\end{abstract}


\end{frontmatter}

\begin{figure}[btp]
\begin{center}\leavevmode
\includegraphics[width=1.0\linewidth]{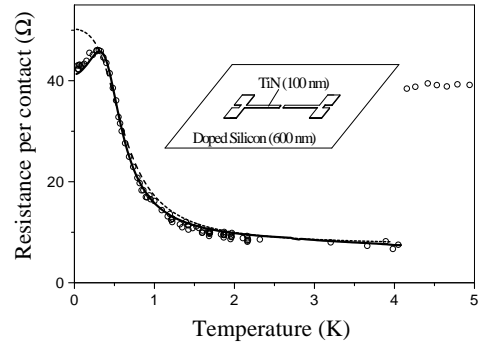}
\caption{Resistance versus Temperature.}\label{fig:R(T)}\end{center}\end{figure}

We present experimental evidence of coherent Andreev backscattering (``Reflectionless tunneling'') \cite{kastal91} in a Superconductor/Semiconductor transmittive junction. The junction is a realization of a SIN contact where TiN is a superconductor ($T_{c}=4.6\,K$) \cite{lef99}, the disordered normal metal a $0.6\,\mu m$ thick layer of highly doped Silicon ($n_{e}=2\,10^{19}\,cm^{-3}$) and I stands for the Schottky barrier. Figure \ref{fig:R(T)} shows the resistance per contact of a $L=20 \,\mu m$ long (distance between electrodes) and $W=10 \,\mu m$ wide sample as a function of temperature.
 
As the temperature is lowered, the resistance drops at the critical temperature of the TiN and increases down to $300\,mK$ due to the reduction of the number of quasiparticules in the TiN electrodes and to imperfect Andreev reflection (Schottky barrier). This behavior is well fitted within the BTK model \cite{BTK} (dashed line in fig:\ref{fig:R(T)}) which gives a transparency per channel $\Gamma \simeq 4\,10^{-2}$, a gap $\Delta=0.22\,mV$ and a damping factor $\Gamma_{s}=0.13\,\Delta$. The value of $\Gamma$ gives a barrier resistance $R_{b}={h\over 2e^{2}}{(\lambda_{F}/2)^2\over \Gamma}\approx 20\,\Omega\,\mu m^{2}$, where $\lambda_{F}$ is the Fermi wavelength in the doped silicon. From the value of the normal resistance of the junction ($R_{NN}=7.5\,\Omega$), we estimate the sheet resistance of the doped silicon {\it underneath} the TiN electrodes : $R_{sheet}^{N}={R_{NN}^{2} W^{2}\over R_{b}}\simeq 320\, \Omega$. This value differs from the bulk value obtained by measurements of the total resistance between very large superconducting pads separated by various distances (Transverse Length Method) $R_{sheet}^{N}\simeq 24\,\Omega$. This difference may be due to some disorder induced by annealing during the process. For temperature $T<300\,mK$, the resistance decreases and the BTK description fails. Introducing the proximity effect \cite{volkov93} we reproduce the observed behavior (solid line in fig:\ref{fig:R(T)}) by adjusting the depairing rate $\simeq (10\,ps)^{-1}$ and {\it keeping} $\Delta$, $\Gamma_{S}$ and $R_{b}$ unchanged. Again, this value differs from the bulk weak localization measurement ($\simeq (2\,ns)^{-1}$).
\begin{figure}[btp]
\begin{center}\leavevmode
\includegraphics[width=0.9\linewidth]{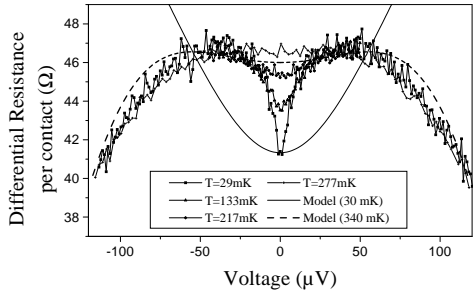}
\caption{Resistance versus applied voltage}\label{fig:R(V)}\end{center}\end{figure}
\par
\begin{figure}[btp]
\begin{center}\leavevmode
\includegraphics[width=1.0\linewidth]{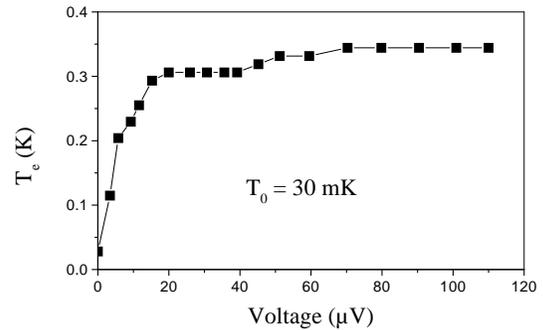}
\caption{Effective carriers temperature versus voltage at $T=30\,mK$}\label{fig:Te(V)}\end{center}\end{figure}

 At low temperature, we have measured the voltage dependance of the differential resistance (fig:\ref{fig:R(V)}). We clearly observe a Zero Bias Anomaly with a maximum of the differential resistance around $40\, \mu V$ which disappears for temperature above $300\,mK$. The shape at high voltage (above the maximum) can be fitted within the BTK model and confirms the large damping factor. Although the maximum appears for an expected voltage ($eV\approx k_{B}T_{peak}$), a more quantitative analysis needs a refined theory. The comparison with the Volkov model is not straightforward at low temperature (solid line in fig:\ref{fig:R(V)}) and we have to introduce an effective temperature $T_{e}$ for the carriers, as a function of the voltage across the barrier (Fig:\ref{fig:Te(V)} at $T_{0}=30\,mK$). We find that $T_{e}$ increases very rapidly at small voltage and saturates around $340\,mK$. This rapid raise is due to the fact that at such low temperature the electron-phonon length is much larger than the distance between the superconducting contacts and that there exists a strong Andreev thermal resistance at the junction interface. At higher temperature (dashed line in fig:\ref{fig:R(V)}), due to the exponential decrease of the Andreev thermal resistance with temperature, the carriers are thermalized with the substrate ($T_{e}=T_{0}$) and the theoretical prediction is close to the observed behavior \cite{strunk}. 
\par We want to thank S. Deleonibus from the LETI/CEA/Grenoble (PLATO) for providing us the TiN/Doped Silicon bilayer.


\end{document}